\documentclass[a4paper]{article}

\usepackage[utf8]{inputenc}
\usepackage{microtype}
\usepackage{erk}
\usepackage{amsmath}
\usepackage{amssymb}
\usepackage{tikz}
\usetikzlibrary{positioning, arrows.meta, shapes, backgrounds}
\usepackage{times}
\usepackage{graphicx}
\usepackage[top=22.5mm, bottom=22.5mm, left=22.5mm, right=22.5mm]{geometry}
\PassOptionsToPackage{hyphens}{url}
\usepackage[hidelinks]{hyperref}
\usepackage{cleveref}
\usepackage[english]{babel}
\usepackage{acro}
\usepackage[export]{adjustbox}
\usepackage{booktabs}
\usepackage{caption}
\usepackage{subcaption}

\DeclareAcronym{ann}{ 
    short = {ANN}, 
    long  = {artificial neural network}
}

\DeclareAcronym{nn}{
    short = {NN},
    long = {neural network}
}

\DeclareAcronym{uv}{
    short = {UV},
    long = {ultraviolet}
}

\newcommand{\etal}{et al.}

\def\footnotemark{}

\begin{document}

\title{Using Modular Arithmetic Optimized Neural Networks To
Crack Affine Cryptographic Schemes Efficiently}

\author{Vanja Stojanovi\'c$^{1}$, Žiga Lesar$^{2}$, Ciril Bohak$^{2}$}

\affiliation{$^{1}$Faculty of Mathematics and Physics, University of Ljubljana, Slovenia\\
E-mail: vs66277@student.uni-lj.si\\
$^{2}$Faculty of Computer and Information Science, University of Ljubljana, Slovenia\\
E-mail: \{ziga.lesar, ciril.bohak\}@fri.uni-lj.si}

\maketitle

\begin{abstract}{Abstract}
We investigate the cryptanalysis of affine ciphers using a hybrid neural network architecture that combines modular arithmetic-aware and statistical feature-based learning. Inspired by recent advances in interpretable neural networks for modular arithmetic and neural cryptanalysis of classical ciphers, our approach integrates a modular branch that processes raw ciphertext sequences and a statistical branch that leverages letter frequency features. Experiments on datasets derived from natural English text demonstrate that the hybrid model attains high key recovery accuracy for short and moderate ciphertexts, outperforming purely statistical approaches for the affine cipher. However, performance degrades for very long ciphertexts, highlighting challenges in model generalization.
\end{abstract}

\selectlanguage{english}

\section{Introduction}
\label{sec:intro}

The cryptanalysis of classical ciphers has long served as a proving ground for both cryptographic and machine learning techniques. Advances in the field have demonstrated that \acp{ann} can be trained to automate attacks on classical ciphers by exploiting statistical features of ciphertexts, such as letter frequencies and n-grams~\cite{Focardi2018}. However, these approaches typically treat the neural network as a black box, without explicitly encoding the algebraic structure underlying many cryptographic schemes.

In parallel, Gromov~\cite{Gromov2023} has shown that simple neural networks can not only learn modular arithmetic operations, but do so in an interpretable and analytically tractable way. In particular, Gromov demonstrates that two-layer networks can ``grok'' modular arithmetic, suddenly generalizing after a period of overfitting, and that the learned weights correspond to periodic, Fourier-like feature maps. This suggests that neural networks can be designed or regularized to explicitly capture the modular structure at the heart of many ciphers, including the affine cipher.

The affine cipher, defined by the transformation \(y = (ax + b) \mod m\), combines modular arithmetic with statistical properties of natural language, making it an ideal testing ground for hybrid approaches. While prior work has leveraged either statistical or algebraic cues in isolation, it remains an open question whether a neural network architecture or training regime that combines explicit modular arithmetic structure with statistical feature learning can improve cryptanalysis of affine ciphers. In this work, we investigate whether integrating modular ari\-thme\-tic-aware neural architectures (as in Gromov~\cite{Gromov2023}) with statistical feature-based learning (as in Focardi and Luccio~\cite{Focardi2018}) can enhance the efficiency and interpretability of neural cryptanalysis for affine ciphers. We analyze not only performance, but also the correlation between algebraic and statistical learning in this context. All code and datasets for this work are available in a GitHub repository.\footnote{\url{https://github.com/Vanja-S/using-modular-arithmetic-optimized-neural-networks-to-crack-affine-cryptographic-shemes-efficiently}}

\section{Related Work}
\label{sec:related-work}

Gromov~\cite{Gromov2023} demonstrates that simple two-layer \acp{ann} can learn modular arithmetic tasks through a phenomenon known as ``grokking,'' where generalization emerges suddenly after extensive training. Notably, the study shows that this learning process corresponds to the discovery of interpretable periodic features, akin to Fourier components. The paper even derives analytic solutions for network weights when learning additive modular functions, as exemplified in \Cref{additive_modular}.
\begin{equation}
\label{additive_modular}
f(n,m) = f_1(n) + f_2(m) \mod p
\end{equation}
This provides strong support for our work by confirming that neural networks can learn the fundamental operations involved in affine ciphers and by offering insights into the mechanism underpinning this learning—namely, the encoding of modular features.

Another relevant contribution explores the use of standard \acp{ann} for automating the cryptanalysis of classical ciphers, including Caesar (shift), Vigenère, and substitution ciphers~\cite{Focardi2018}. This study adopts a ciphertext-only setting and leverages known statistical weaknesses of the ciphers. For the shift cipher, a neural network is trained to map ciphertext frequency distributions to the corresponding shift key. This network is then repurposed to attack Vigenère ciphers by segmenting ciphertext into monoalphabetic subtexts based on hypothesized key lengths and applying the trained shift model. This approach validates the broader applicability of neural networks in classical cryptanalysis, reinforcing the feasibility of our method for the affine cipher.

Further work by Jeong~\etal~\cite{Jeong2024} expands on the potential of deep learning for cryptanalysis. Their findings highlight how neural models can effectively reveal vulnerabilities in encryption schemes, illustrating the capacity of neural cryptanalysis to discover weaknesses in cryptographic algorithms. This aligns closely with Gromov's results, affirming neural networks as powerful tools for addressing cryptographic challenges.

Finally, Dubey~\etal~\cite{Dubey2021} focus on incorporating modular arithmetic into neural network design. Their research emphasizes the development of architectures that naturally accommodate the structural requirements of cryptographic tasks. These insights directly inform our approach, which integrates modular branches within the network to enhance its cryptanalytic capabilities.

\section{Data Generation, \ac{ann} and Cipher Implementation}
\label{sec:method}

To evaluate our neural network's architecture (proposed in \Cref{subsec:ANN}) for the affine cipher cryptanalysis, we constructed a dataset of plaintext-ciphertext pairs with corresponding encryption keys. The data generation process was designed to ensure both diversity and reproducibility, and to reflect realistic cryptanalytic scenarios.

Plaintext samples were taken from the Project Gutenberg English language corpus, which provides a large and varied collection of natural language text. All text was preprocessed by removing punctuation, converting to uppercase, and mapping each character to its corresponding integer value in the range 0 to 25 with \Cref{mapping}.
\begin{equation}
    \label{mapping}
    h: \{A,B,C,\dots,Z\} \to \mathbb Z_{26}
\end{equation}
i.e., \(A \mapsto 0, B \mapsto 1, \ldots, Z \mapsto 25\). Non-alphabetic characters were discarded to maintain consistency with the affine cipher’s domain.

For each plaintext sample, a distinct affine cipher key was generated at random. The key comprises two integers, \(a\) and \(b\), where \(a\) is drawn uniformly at random from the set of non-zero elements of $\mathbb Z_{26}$ that are coprime to the alphabet size (that is \(m = 26\)), and \(b\) is drawn uniformly at random from \(\{0, 1, \ldots, 25\}\). There are exactly 12 possible values for \(a\), corresponding to the integers in this range for which \(\gcd(a, 26) = 1\). The requirement that \(a\) be coprime to 26 ensures that the encryption function is invertible, which is necessary for the affine cipher to be valid.

The affine cipher encrypts each plaintext letter \(x\) according to the transformation \Cref{affine_cipher}.
\begin{equation}
    \label{affine_cipher}
    y = (ax + b) \mod m
\end{equation}
where \(y\) is the ciphertext letter, and all operations are performed modulo \(m = 26\). This transformation was implemented in Python, and applied to each plaintext sample using its associated key. The resulting ciphertexts, along with their corresponding keys and plaintexts, were stored for subsequent use in model training and evaluation.

The final dataset consists of tuples \((C, K, P)\), where \(C\) is the ciphertext sequence, \(K = (a, b)\) is the key, and \(P\) is the original plaintext sequence. All sequences were padded or truncated to a fixed length \(L\) to facilitate batch processing in neural network training. The dataset was randomly shuffled and split into training, validation, and test sets in proportions of 80\%, 10\%, and 10\%, respectively.

All data processing and encryption routines were implemented in Python 3.11, leveraging the NumPy and PyTorch libraries for efficient tensor operations. All models were also implemented with PyTorch. The dataset was stored in PyTorch tensor format, with each batch containing ciphertexts, keys, and plaintexts as separate tensors.

\subsection{Modular Arithmetic-Aware Neural Network Architecture}
\label{subsec:ANN}

To effectively exploit the algebraic structure of the affine cipher, we designed a hybrid neural network architecture that processes both the raw ciphertext sequence and statistical features derived from the ciphertext. This design is inspired by the analytic solutions described by Gromov~\cite{Gromov2023} and by the statistical feature-based approach by Focardi and Luccio~\cite{Focardi2018}. The goal is to enable the network to learn both modular arithmetic patterns and language statistics relevant for cryptanalysis.

\paragraph{InputRepresentation:} Each input sample consists of:
\begin{itemize}
    \item A ciphertext sequence \(C = (c_1, c_2, \dots , c_L)\), where each \(c_i \in \mathbb Z_{26}\), represented as a vector of integers of length \(L\), where L is the size of the dataset \(L \in \{100, 500, 1000, 10\,000\}\).
    \item A statistical feature vector \(S \in 	\mathbb{R}^{26}\), representing the normalized frequency of each letter in the ciphertext.
\end{itemize}

\paragraph{Network Architecture} The network consists of two parallel branches:

\vspace*{0.1cm}
\noindent
\textbf{Modular Branch:}
\begin{itemize}
    \item \textit{Embedding Layer}: Maps each integer \(c_i\) to a 16-dimensional learnable vector, producing an embedding matrix of shape \(L \times 16\).
    \item \textit{Flattening}: The embedding matrix is flattened into a single vector of length \(16L\).
    \item \textit{Dense Layer 1}: A fully connected layer with ReLU activation, mapping the flattened vector to a hidden dimension (128 in our experiments).
    \item \textit{Dense Layer 2}: Another fully connected layer with ReLU activation, producing the modular feature vector (dimension 128).
\end{itemize}

\noindent
\textbf{Statistical Branch:}
\begin{itemize}
    \item \textit{Dense Layer 1}: A fully connected layer with ReLU activation, mapping the 26-dimensional frequency vector to a hidden dimension (128 in our experiments).
    \item \textit{Dense Layer 2}: Another fully connected layer with ReLU activation, producing the statistical feature vector (dimension 128).
\end{itemize}

\begin{figure*}[ht!]
    \centering
    \vspace{-0.5cm}
    \begin{adjustbox}{max width=0.75\textwidth}
        \begin{tikzpicture}[node distance=1cm, auto, >=latex, thick, every node/.style={transform shape}]
            \node[draw, rectangle, minimum width=2.5cm, minimum height=0.75cm, fill=blue!10] (cipher) {Ciphertext Sequence};
            \node[draw, rectangle, right=0.9cm of cipher, minimum width=2.5cm, minimum height=0.75cm, fill=blue!20] (embed) {Embedding Layer};
            \node[above=0.25cm of embed] (modular) {\Large Modular branch};
            \node[draw, rectangle, right=0.9cm of embed, minimum width=1.4cm, minimum height=0.75cm, fill=blue!30] (flat) {Flatten};
            \node[draw, rectangle, right=0.9cm of flat, minimum width=2.5cm, minimum height=0.75cm, fill=blue!40] (dense1) {Dense + ReLU};
            \node[draw, rectangle, right=0.9cm of dense1, minimum width=2.5cm, minimum height=0.75cm, fill=blue!50] (dense2) {Dense + ReLU};

            \node[draw, rectangle, below=1.25cm of embed, minimum width=2.5cm, minimum height=0.75cm, fill=red!10] (stat) {Letter Freq. Vector};
            \node[above=0.25cm of stat] (stat1) {\Large Statistical branch};
            \node[draw, rectangle, right=0.9cm of stat, minimum width=2.5cm, minimum height=0.75cm, fill=red!20] (sdense1) {Dense + ReLU};
            \node[draw, rectangle, right=0.9cm of sdense1, minimum width=2.5cm, minimum height=0.75cm, fill=red!30] (sdense2) {Dense + ReLU};

            \node[draw, rectangle, right=0.9cm of sdense2, minimum width=2.5cm, minimum height=0.75cm, fill=green!40] (final) {Final Dense Layer (312 logits)};
            \node[right=0.9cm of final] (output) {Predicted Key $(a, b)$};

            \draw[->] (cipher) -- (embed);
            \draw[->] (embed) -- (flat);
            \draw[->] (flat) -- (dense1);
            \draw[->] (dense1) -- (dense2);
            \draw[->] (dense2) -- (final);

            \draw[->] (cipher) -- (stat.west);
            \draw[->] (stat) -- (sdense1);
            \draw[->] (sdense1) -- (sdense2);
            \draw[->] (sdense2) -- (final);

            \draw[->] (final) -- (output);
        \end{tikzpicture}
    \end{adjustbox}
    \caption{Neural network architecture for affine cipher key recovery.}
    \label{fig:diagram}
    \vspace{-.5cm}
\end{figure*}
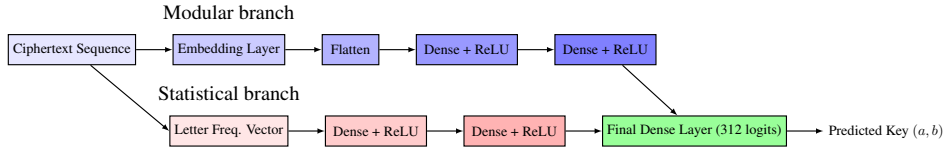

The outputs of both branches (each of size equal to the hidden dimension) are concatenated and passed through a final fully connected layer \emph{without activation}, which outputs a vector of logits of length 312, corresponding to all possible affine keys (\(a, b\)). These logits are used as input to the cross-entropy loss function during training.

\paragraph{System diagram:} A diagram of the ANN architecture is presented in \Cref{fig:diagram}.

\paragraph{Training Objective:} The network is trained to minimize the cross-entropy loss between the predicted logits and the true key class index for each sample. The output logits are interpreted as unnormalized scores for each possible affine key (\(a,b\)), and the predicted key is the one with the highest logit. The model is trained end-to-end using the Adam optimizer and standard backpropagation.

\paragraph{Implementation Details:} Hyperparameters, including the number of layers, hidden units (128), and activation functions (ReLU), were selected based on initial validation and kept fixed for all experiments. The modular and statistical branches are trained jointly end-to-end. While the modular branch does not explicitly encode modular arithmetic, its design --- processing the raw ciphertext sequence via embeddings --- enables the network to learn modular patterns from data.

\section{Results and Analysis}
\label{sec:results}

\subsection{Experimental results}
We trained the proposed ANN on the affine cipher key recovery task using ciphertexts of varying lengths (\(L\) = 100, 500, 1000, 10\,000). The model was trained for 30 epochs with a hidden layer size of 128 and a batch size of 128. The results are summarized in \Cref{tab:results}.

\begin{table}[h]
    \centering
    \small
    \begin{tabular}{ccccc}
        \toprule
        Length & Hidden & Batch & Test Acc. (\%) & Epochs \\
        \midrule
        100   & 128 & 128 & 98.08 & 30 \\
        500   & 128 & 128 & 96.85 & 30 \\
        1000  & 128 & 128 & 70.53 & 30 \\
        10000  & 128 & 128 & 2.50 & 30 \\
        \bottomrule
    \end{tabular}
    \caption{Test accuracy for affine key recovery as a function of ciphertext length.}
    \label{tab:results}
\end{table}

\begin{figure*}
    \centering
    \begin{subfigure}[b]{0.44\textwidth}
        \centering
        \includegraphics[width=\textwidth]{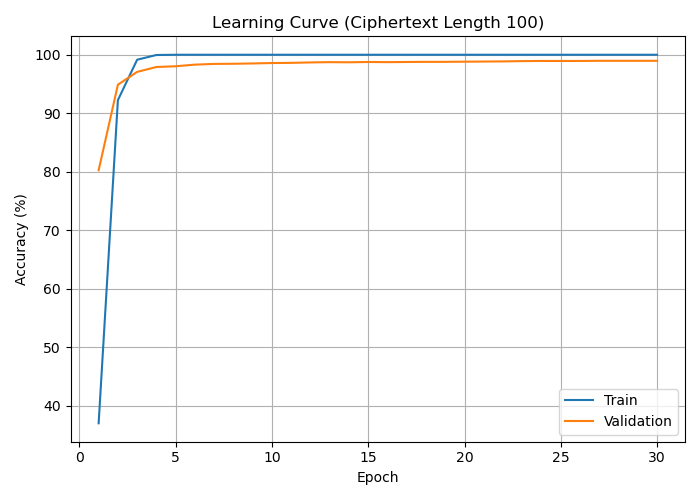}
        \caption{Ciphertext length is 100.}
        \label{fig:learning-curve-1}
    \end{subfigure}
    \hspace{1.05cm}
    \begin{subfigure}[b]{0.44\textwidth}
        \centering
        \includegraphics[width=\textwidth]{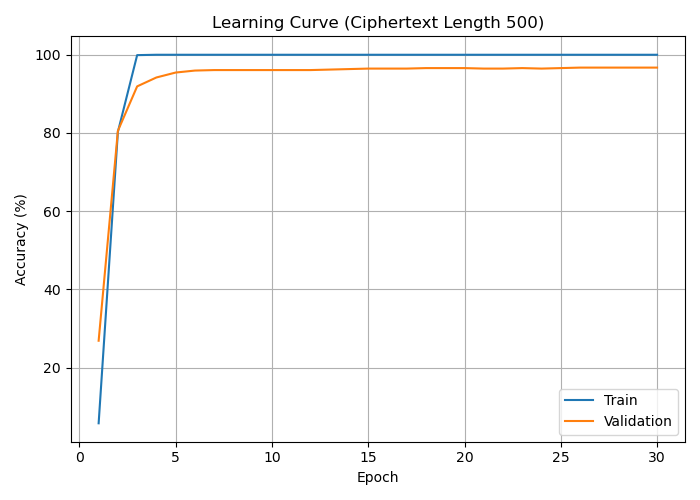}
        \caption{Ciphertext length is 500.}
        \label{fig:learning-curve-2}
    \end{subfigure}
    \\
    \begin{subfigure}[b]{0.44\textwidth}
        \centering
        \includegraphics[width=\textwidth]{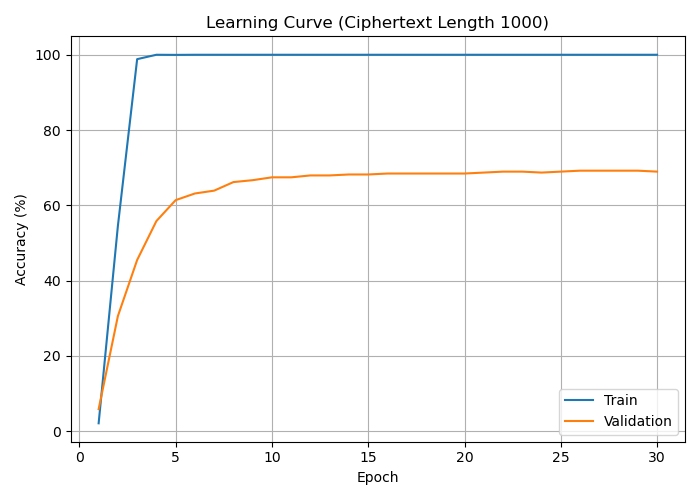}
        \caption{Ciphertext length is 1\,000.}
        \label{fig:learning-curve-3}
    \end{subfigure}
    \hspace{1.05cm}
    \begin{subfigure}[b]{0.44\textwidth}
        \centering
        \includegraphics[width=\textwidth]{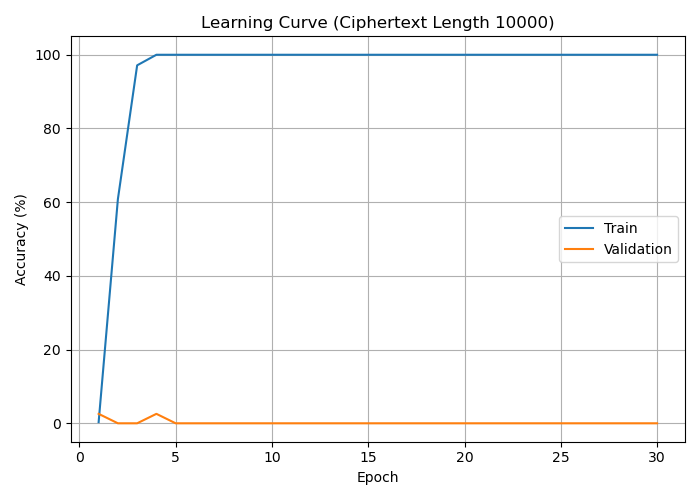}
        \caption{Ciphertext length is 10\,000.}
        \label{fig:learning-curve-4}
    \end{subfigure}

    \caption{Learning curve (train and validation accuracy) for ciphertexts of different lengths.}
    \label{fig:three graphs}
    \vspace{-0.5cm}
\end{figure*}

\begin{figure}[ht!]
    \centering
    \includegraphics[width=0.44\textwidth]{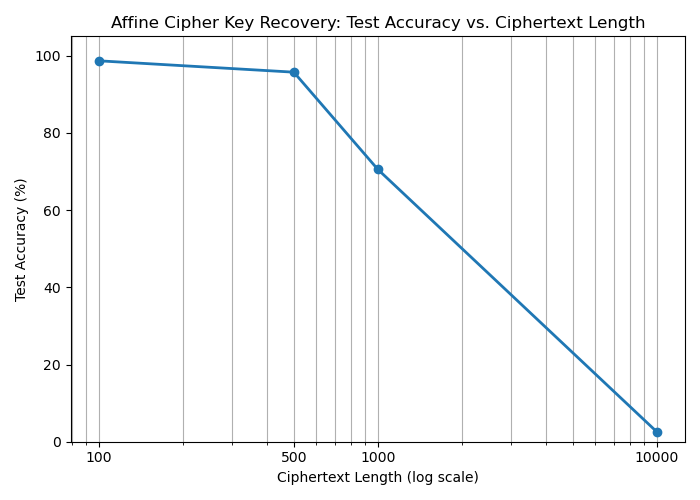}
    \caption{Test accuracy for affine key recovery as a function of ciphertext length.}
    \label{fig:acc}
    \vspace{-0.5cm}
\end{figure}

The learning curves for each ciphertext length are sh\-own in \Cref{fig:three graphs,fig:acc}. For \(L = 100\) and \(L = 500\), the model achieves near-perfect accuracy after only a few epochs, with both training and validation accuracy converging ra\-pi\-dly. For \(L = 1\,000\), the model also achieves perfect training accuracy, but the validation and test accuracy plateau at a lower value, suggesting overfitting or a limitation in the model’s ability to generalize for longer ciphertexts under the current setup. The same goes for the longer text of \(L = 10\,000\) where the testing accuracy is significantly lower and the validation confirms that the neural network is not able to predict the key.

Additionally, \Cref{fig:acc} shows the test accuracy as a function of ciphertext length, summarizing the model’s performance across all settings.

\subsection{Discussion}

The results demonstrate that the hybrid neural network is highly effective at recovering the affine cipher keys from ciphertexts of moderate length (\(L = 100\) and \(L = 500 \)), achieving test accuracies above 96 \%. The learning curves indicate rapid convergence and strong generalization for these settings. For longer ciphertexts (\(L = 10\,000\)), the model achieves perfect training accuracy but lower test accuracy, suggesting that either the model capacity, the feature representation, or the training regime may need to be further tuned to handle longer sequences without overfitting.

The observed performance drop for \(L = 10\,000\) may be due to the increased complexity of the input, the fixed model size, underfitting, the embedding layer pro\-p\-er\-ties of the modular branch, or the statistical properties of the dataset. Further investigation, such as increasing the mod\-el capacity, using regularization, or augmenting the feature set, could help address this limitation.

\subsection{Comparison with Forcardi and Luccio}

Focardi and Luccio~\cite{Focardi2018} demonstrated that standard artificial neural networks can be trained to recover keys for classical ciphers, including the Caesar and Vigenère ciphers, by leveraging statistical features such as letter frequencies and \(n\)-grams. Their approach achieved high accuracy for short, mono-alphabetic ciphers, but did not explicitly incorporate the algebraic structure of the cipher into the neural network architecture.

In contrast, our approach combines both modular ari\-thmetic-aware and statistical feature-based learning in a hybrid neural network, inspired by the analytic insights of Gromov~\cite{Gromov2023}. Our results show that this hybrid model can achieve comparable or superior accuracy for affine ciphers, particularly for shorter ciphertext lengths (under 500 in length). The rapid convergence and high accuracy for \(L = 100\) and \(L = 500\) suggest that explicitly encoding modular structure, in addition to statistical features, provides a significant advantage for cryptanalysis of ciphers with modular arithmetic components.

Moreover, our experiments highlight the importance of model design and feature selection in neural cryptanalysis. While Focardi and Luccio’s method is effective for ciphers where statistical features dominate, our results indicate that hybrid models are better suited for ciphers like the affine cipher, where both algebraic and statistical properties are essential for successful key recovery.

\subsection{Limitations and Future Work}
While the hybrid model performs well for moderate ciphertext lengths, its performance degrades for longer sequences. Future work should explore increasing model capacity, incorporating additional regularization, and experimenting with alternative architectures (such as transformers or convolutional networks) to improve generalization. Additionally, further ablation studies could clarify the relative contributions of the modular and statistical branches, and experiments on other modular ciphers (e.g., Hill cipher) could extend the generality of these findings.

\section{Conclusion}
\label{sec:conclusion}

We have shown that the proposed hybrid neural network combining modular arithmetic-aware and statistical feature-based le\-ar\-ning can accurately recover affine cipher keys from mo\-de\-ra\-te-length ciphertexts. Explicitly modeling modular structure enables superior performance over purely statistical approaches for shorter texts. However, the model’s effectiveness diminishes with longer ciphertexts, indicating challenges in generalization. These results emphasize the value of incorporating algebraic priors into neural cryptanalysis, especially for ciphers with modular components. Future work should focus on improving scalability and robustness, as well as extending this approach to other cryptographic schemes. Our findings highlight the promise of interpretable, structure-aware neural models for advancing automated cryptanalysis.

\footnotesize
\bibliographystyle{plain}
\bibliography{bibliography}

\end{document}